\documentclass[12pt,twoside]{article}
\usepackage{fleqn,espcrc1}
\usepackage{graphicx}

\newcommand {\GOTO} {$\rightarrow$}

\newcommand {\XF}   {x$_F$}
\newcommand {\PT}   {p$_T$}

\newcommand {\DLL}   {D$_{NN}$}
\newcommand {\DNN}   {D$_{NN}$}

\newcommand {\LA}   {$\Lambda$}

\newcommand {\SI}   {$\Sigma$}
\newcommand {\XI}   {$\Xi$}

\newcommand {\XIM}  {$\Xi^-$}

\newcommand {\SIO}  {$\Sigma^0$}
\newcommand {\SIP}  {$\Sigma^+$}
\newcommand {\SIM}  {$\Sigma^-$}
\newcommand {\KM}   {K$^-$~}
\newcommand {\KP}   {K$^+$}

\newcommand {\PI}   {$\pi$}
\newcommand {\PIO}  {$\pi^0$}
\newcommand {\PIM}  {$\pi^-$}
\newcommand {\PIP}  {$\pi^+$}

\newcommand {\XIC}  {$\Xi_c$}

\newcommand {\PBAR} {$\overline{p}$}

\title{Strangeness in Hadronic Interactions}

\author{Stephan~Paul\\
Technische Universit\"{a}t M\"{u}nchen,\\Physik-Department,\\
James-Franck Stra\ss e, D-85748 Garching, Germany}

\begin{document}

\maketitle

\begin{abstract}
Strangeness has always been an important subject at all PANIC
conferences as it probably constitutes the best link between
particle and nuclear physics. I will thus use the theme of the
conference by considering strangeness as a tourist through the
world of strong interaction. During this talk we will accompany
strangeness from production, to the royaume of mesons and baryons
up to the complex world of nuclei.
\end{abstract}
\section{Production of and by strange particles}
\subsection{Production of hyperons by hyperons at high energies}
\begin{figure}[htbp]
\vspace{-7.5 cm} \hspace{7.5cm}
\begin{minipage}[htpb]{8.5cm}
  \begin{center}
    \includegraphics[width=6cm,draft=false]
    {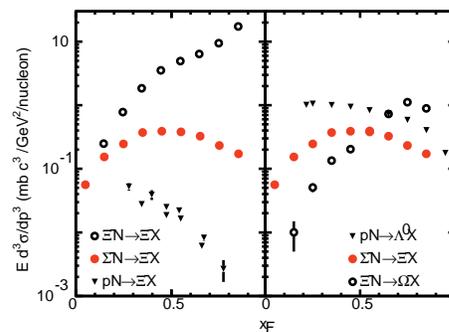}
   \end{center}
   \vspace{-2.4 cm}

   \caption{\small Hyperon production cross sections by WA89}
   \label{hyperon-production}
\end{minipage}
\end{figure}

The production characteristics of hyperons at high energies has
since long been studied. Using recent results from the hyperon
beam at CERN \cite{wa89_production} we now have a systematic set of data on
differential cross sections for the production of different
hyperons using beams with different strangeness (S =0,-1,-2).The
resulting systematic is shown in fig \ref{hyperon-production}. It
shows that different production processes are at play for central
production (\XF $<$ 0.3) and forward production (\XF $>$ 0.7). At
central collisions the strange quark for the final hyperon is
produced from the sea and little difference is seen if strangeness
is already brought along in the beam. As is well known, the cross
section drops by a factor 10 for

\noindent
\begin{minipage}[htpb]{6.5cm}
each unit of strangeness produced without exhibiting any
exhausting behavior coming from a reduction in cms-energy once 1
or 2 strange quarks have already been produced. The situation is
drastically different at forward kinematics, where the final
hyperon profits from a strange quark in the beam. Each additional
unit of strangeness again leads to a punishment, but now in the
order of a factor 100, probably due to the limited energy
available in the initial collision.
\end{minipage}

\subsection{Production of charm states by hyperons}
In the past it had been speculated that hyperons as projectile
could exhibit a larger cross section for charm production that
protons or even mesons. Even in the absence of exotic reasons one
expected the harder momentum distribution of the strange quark in
the projectile to boost the cross section in the kinematic region
where it strongly depends on the beam energy. The measured cross section
using charmed and charmed strange mesons as well as baryons
($\sigma_{tot}$ = 11.3$\pm$2.4$\pm$2.0 $\mu$barn)
\cite{WA89_charm} perfectly agrees with existing proton data and
their predicted kinematical dependence.

As inferred from the hyperon production data the distribution for
final states is altered and an efficient production of charmed
strange systems is observed (as well as baryons over mesons)
resulting in a strong asymmetry for D$_s^-$/D$_s^+$ in the forward
region \cite{WA89_asym} .

\subsection{Polarization of hyperons in hadronic interactions - The case of \DLL~at low
energies}
Polarization of hyperons have been observed since many
years. Mostly these measurements were done at high energies (many
GeV beams) and different final states were studied. No complete
picture of the phenomena observed yet exists. However, besides of
polarization production one can also measure polarization transfer
from a transversely polarized beam to a hyperon, usually called
\DNN, the depolarization coefficient in the normal to the
production plane. While at high energy the phenomena must be
explained in a quark picture the low energy regime is the
playground for meson exchange pictures concurring with
descriptions on quark exchange. While threshold production in
\PBAR p annihilation has already been reported some time ago I
would like to mention new results from Saturne. Using a
transversely polarized proton beam they observed Lambda
polarization in exclusive processes. The different hyperons being
produced primordially could be distinguished by final state
analysis and missing mass techniques. The results of the DISTO
experiment show a negative polarization transfer from the proton
to the Lambda of about -20\% - -50\%, increasing with \XF~and decreasing

\begin{figure}[htbp]
\vspace{-6.5 cm} \hspace{8.3 cm}
\begin{minipage}[htpb]{7.5 cm}
  \begin{center}
    \includegraphics[width=6cm,draft=false]
    {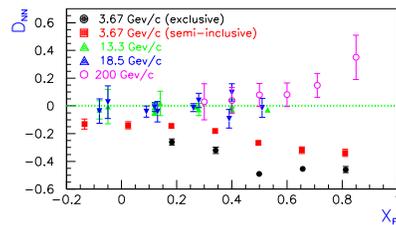}
   \end{center}
   \vspace{-1.6 cm}
   \caption{\small $\Lambda$-polarization transfer from DISTO}
   \label{DISTO}
\end{minipage}
\end{figure}

\noindent
\begin{minipage}[htpb]{7.5cm}
with \PT~\cite{DISTO}(see fig.\ref{DISTO}). This can well be
explained by a \KM-meson exchange which leads to the production of
strange hadrons in the final state (pion exchange on the other
hand would predict a large and positive value for \DNN) However,
if we consider the same process at higher energies the effect
shows a change in sign. Clearly, the data at 200~GeV are inclusive
production data but that should not lead to a change in sign.
\end{minipage}

\subsection{Polarization of strange nuclei}
Polarized hypernuclei are an useful tool to improve on the
spectroscopy of hypernuclei, determine the magnetic moment of \LA~
in the nuclear medium and to further study the dynamics of mesonic
and non-mesonic decays of \LA-hypernuclei. \LA-polarization in low
energy exclusive reactions can be calculated from an phase shift
analysis of K and \PI~scattering involving hyperons in the final
state. In a typical hypernuclear experiment n(\PIP,\KP)\LA~the
\LA-production can be explained by phase shift analysis of \PI-nucleon
scattering using Regge exchange and assuming intermediate baryon
resonances. The underlying free process has shown to give rather
large polarization of free Lambdas (60-80\%). Lambda Polarization
in nuclei using the (\PIP,\KP) reaction has first been observed by
\cite{ajimura1} using the quasi-free Lambda production in $^{12}$C
(P$_{\Lambda}\sim$ 60-65\%). The results agreed very well with the
calculation assuming the production on free neutrons, corrected
for the Fermi-momentum effect.

These calculations of the free Lambda polarization has to be
adapted for nuclear effects, which can lead to a depolarization by
gamma-emission or particle emission \cite{Motoba}. It
was pointed out that $^5_{\Lambda}$He is an ideal system to study
the polarization and test the model predictions. In a recent
experiment at KEK \cite{ajimura2} this system was produced by the
$^6$Li(\PIP,\KP p)$^5_{\Lambda}$He reaction, a strangeness
exchange reaction on Lithium with subsequent proton emission at
p(\PI) = 1.05 GeV/c. Owing to the low nuclear mass number the
decay of the hypernucleus shows a strong mesonic component with
large decay asymmetry from the free \LA~decay. A polarization of
~25\% (0.247$\pm$0.082) was observed at small angles (transverse
momenta) and of ~40\% (0.393$\pm$0.094) at larger angles. The
results are well reproduced by the calculations giving 18\% and
37\%, respectively. Considering the complex production mechanism
this results gives confidence that the polarization of other
hypernuclei may be extracted from theory and may be used to study
other phenomena (as mentioned above).
\section{Strangeness in  Hadrons (interaction with the host)}
\subsection{The size of strange baryons}
\noindent
\begin{minipage}[htpb]{7.5cm}
The structure of hadrons is closely connected to the question of
their size. We can define hadronic and charge radii for mesons and
baryons, as they can be measured by elastic scattering and
electron scattering \cite{Povh91} and which can be explained in
different models. Here we shall only deal with the more familiar
concept of charge radii. Assuming an effective size of a
constituent quark which scales with the mass of the quark Povh et
al. have predicted the radii for all different hadrons. While most
of the hadronic radii have been determined, electromagnetic radii
for hyperons are still missing. This field has recently seen
important progress. Using the method of electron scattering in
reverse kinematics (scattering charged hyperons off the electrons
in a target) which
\end{minipage}
\begin{figure}[htbp]
\vspace{-10.5 cm} \hspace{8.5cm}
\begin{minipage}[htpb]{7.5cm}
  \begin{center}
    \includegraphics[width=6cm,draft=false]
    {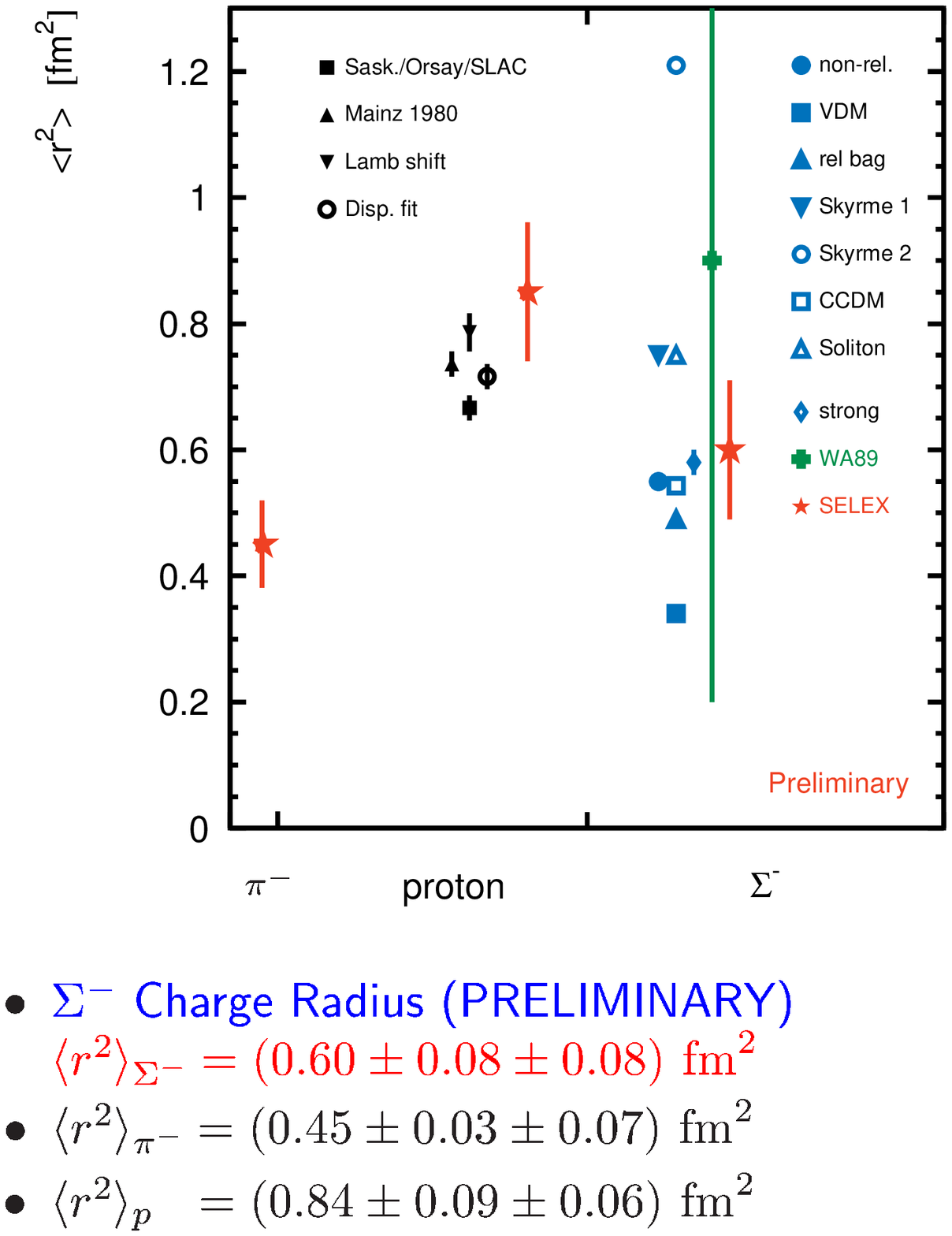}
   \end{center}
   \vspace{-1.25 cm}
   \caption{\small Charge radii of proton, $\pi$ and $\Sigma^-$}
   \label{SELEX_radii}
\end{minipage}
\end{figure}

\noindent has been used for the determination of the \PI~and K
radii data on the charge radius of the \SIM~could be obtained. The
first measurement by WA89 \cite{WA89_electron} establishing the
method which showed large statistical and systematic errors was
recently superseded by a more precise result from the SELEX
\cite{SELEX} group at FNAL. The mean square charge radius is
determined by the slope of the elastic form factor at q$^2$=0. The
SELEX experiment has cross checked their results by a simultaneous
measurement of the pion and proton radii. The results are depicted
in fig.\ref{SELEX_radii} together with a number of predictions.
As for the meson system (\PI-K) the strange counterpart of the
proton shows a smaller radius than the charged nucleon confirming
the predictions of many models. The key measurement to expect in
the future is a measurement of \SIP~ charge radius which should be
larger than the proton one owing to the different charge of u and
d-quarks. Many other models exist and most of them predict the
\SIP~to be smaller than the proton, just as the \SIM.
\subsection{The role of strangeness in meson and baryon spectroscopy}
\noindent
\begin{minipage}[htpb]{8.5cm}
As we know from the constituent quark model and the many more
'realistic' models derived from it later on the binding forces
inside a hadron can be derived from a series of potentials.
A central potential and higher order terms like spin-orbit or
spin-spin interaction. At short distances the interaction of
quarks proceeds analog to e.m. via the 1-gluon exchange.
\end{minipage}

\begin{figure}[htbp]
\vspace{-4.9cm} \hspace{9cm}
\begin{minipage}[htpb]{6cm}
  \begin{center}
    \includegraphics[width=2.2cm,draft=false]
    {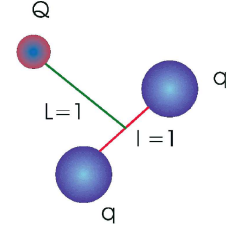}
   \end{center}
   \vspace{-1.3 cm}
   \caption{\small Baryons in the constituent quark model}
   \label{ISGUR}
\end{minipage}
\end{figure}

\vspace{-.7cm}
\noindent
Similarly
we can define a spin-spin splitting where the couplings are assumed to
be proportional to the color magnetic moments of the hadrons.
Since the magnetic moment

\noindent
\begin{minipage}[htpb]{7.6cm}
of a constituent quark scales with its
mass we have an effective flavor breaking of the quark-quark
interaction. Negative parity states show a richer structure
particularly in the baryon sector. In the quark model description
two orthogonal degrees freedom which are defined combining only
two quarks or the axis of the third quark moving around the other
two. In equal mass systems these two systems are degenerate but
introducing a different quark flavor breaks this symmetry and the
system\end{minipage}

\begin{figure}[htbp]
\vspace{-8.25 cm} \hspace{8.5cm}
\begin{minipage}[htpb]{7.5cm}
  \begin{center}
    \includegraphics[width=6cm,draft=false]
    {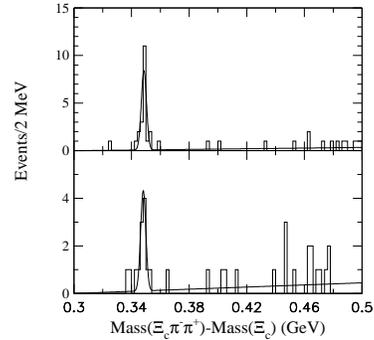}
   \end{center}
   \vspace{-1.25 cm}
   \caption{\small New $\Xi_c$ state observed by CLEO}
   \label{CLEO}
\end{minipage}
\end{figure}

\vspace{-.6cm}
\noindent
splits into \LA~and \SI-type states (symmetric and
anti-symmetric in the exchange of two equal mass quarks)
\cite{Isgur78}. This becomes more interesting if angular momentum
is involved

\noindent
\begin{minipage}[htpb]{7.5cm}
as these two coordinates can be distinguished more
easily (fig.\ref{ISGUR}). While for l=0 the \LA-states are lighter than the \SI~
states the particular symmetry leads to a partial inversion
of this order for l$\geq$1.
For \LA-type baryons we expect the
angular momentum to be mostly within the qq-coordinate, thus
increasing their distance and decreasing the spin-spin force,
while for the \SI-type system the angular momentum is mostly in
the orthogonal coordinate of the quark with unequal mass.
Replacing the strange quark by a charm quark reduces the
$\frac{3}{2}^+-\frac{1}{2}^+$ splitting in the \SI-system
\end{minipage}

\begin{figure}[htbp]
\vspace{-7.7cm} \hspace{8.5 cm}
\begin{minipage}[htpb]{8 cm}
  \begin{center}
    \includegraphics[width=5.9cm,draft=false]
    {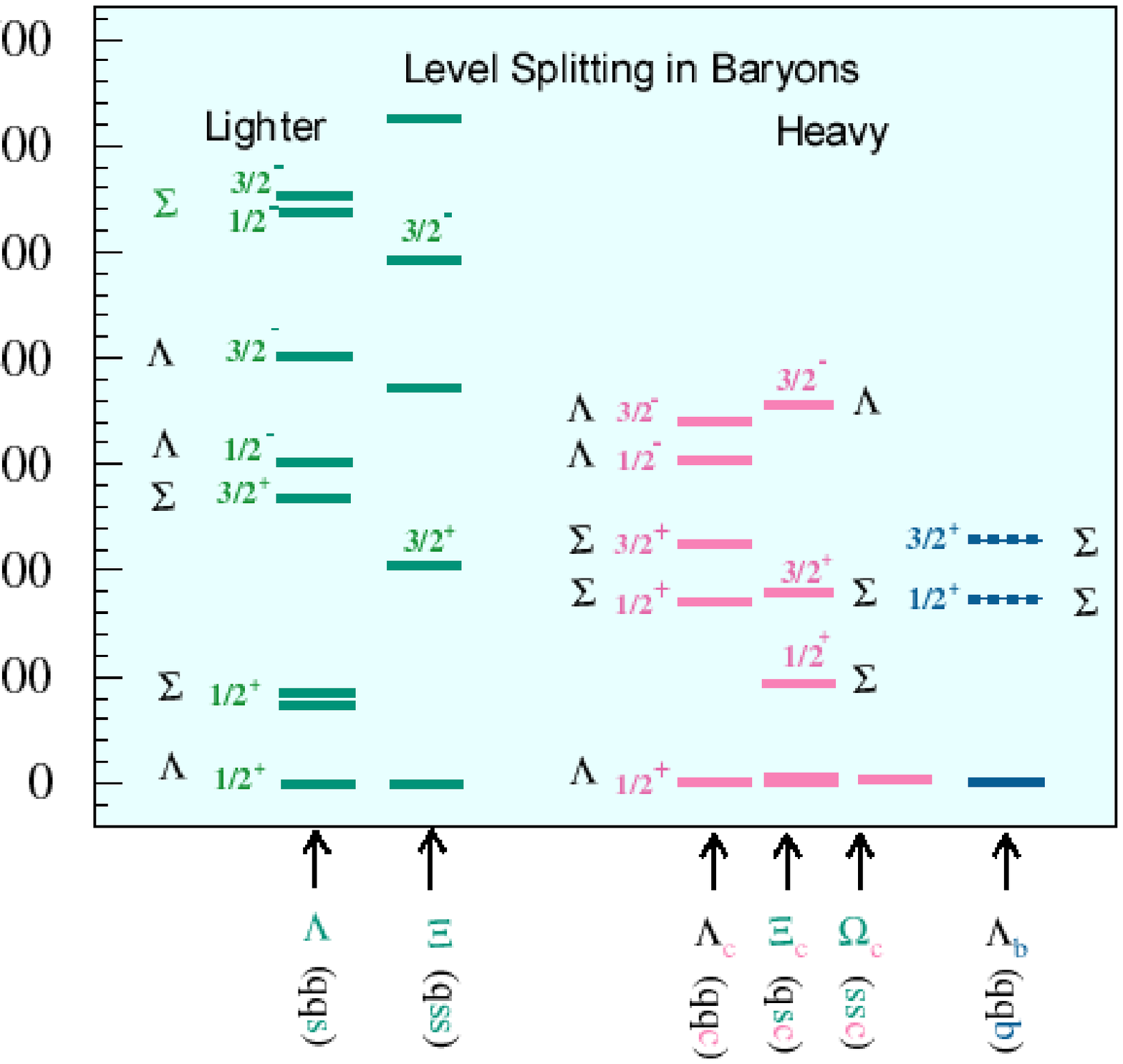}
   \end{center}

   \vspace{-1.5cm}
   \caption{Baryon masses}
   \label{baryon_masses}
\end{minipage}
\end{figure}

\newpage
\noindent
(because of the heavy charm quark mass)
but increases the \LA-\SI~
splitting.
Adding a strange quark to the charm quark almost
reestablishes the \LA-\SI~situation because the repulsion in the
S=1 qs-system is reduced and compensates the lack of spin-spin
interaction of lighter and charm quark.
Also the spin
$\frac{3}{2}^+$ state is moved down for the same reason (see also fig.\ref{baryon_masses}. New
results are now available for the l=1 system, particularly in the
csq-sector. For these states in the csq-sector the spin-spin
interaction is reduced to the minimum for the \LA-type system, as
the two lighter quarks are now at larger distances due to the
angular momentum between them. We expect the \LA~$\frac{1}{2}^-$
and $\frac{3}{2}^-$ state to be almost degenerate, which still
shows a small separation in the qqc system.
Indeed, the \LA $\frac{1}{2}^-$ system of the \XIC~has not yet been found.
CLEO \cite{CLEO} has recently reported the observation of a new
\XIC-state, interpreted as $\frac{3}{2}^-$, decaying via
2\PI-emission into the ground state \XIC~(fig.\ref{CLEO}). The
extreme narrow width of this state and the particular decay
pattern (via an intermediate \XIC$^*$) supports the spin
assignment and the baryon picture drawn above,
using arguments too long to be outlined here.
\subsection{Lifetimes of heavy strange hadrons}
Including strangeness into charmed baryons results in a rich
pattern of lifetime observed for various charmed strange baryons.
This result is caused by an interference of the s-quark stemming
from the c-quark decay and the one present in the parent baryon.
While short distance physics puts lifetimes of all charmed hadrons
to about 1 ps, long range correlations lead to a variety spanning
a factor of 20 with the $\Omega_c$ being the shortest one.

\section{Hypernuclei (Strangeness visiting nuclei)}
The pioneering work in this field was done using recoiless
production of \LA-hypernuclei (\KM,\PIM) at 700-800 MeV/c, which
established the existence of hypernuclei and allowed a clear
assignment of hypernuclear states from where the \LA-N potential
was derived and first estimates on \LA~spin-orbit potential were
obtained. Comparing the features to heavy baryons we can see
similar and very different features. a) While the confinement
potential is flavor independent (and thus the central constituent
quark potential) the \LA-N interaction is weakened as compared
with the NN to about 3/5 leading to a potential depth of 30 MeV as
compared to 50 MeV for the NN case. b) As is the case for heavy
quarks the 'intruder' is hardly sensitive to the spin interaction
(LS and SS) which results in a quasi degeneration of states of the
new system. c) The heavy visitor is well distinguishable from the
ordinary population (light quarks and nucleons) as the many
body system can be described as a core coupling to the 'tourist'
which mainly behaves as an independent single particle system acting
as a static source of attraction.

Lately the development has shifted to the (\PIP,\KP) reaction
pursued at KEK and BNL. Owing to its large recoil (~350 MeV/c)
this reaction allows to populate also the low lying ground states
of hypernuclei. The population favors maximum angular momentum
(spin-stretched states) of the core/\LA~system.

The virtue of this reaction can be seen in the $^{89}_{\Lambda}$Y
spectrum (see fig.\ref{Yttrium}) where the core nucleus mainly stays in a
g$_{9/2}$ state and the excitation spectrum clearly unravels the
single particle spectrum of Lambda's in a nucleus. It has since
been the text book example of a single particle spectrum within a
strongly interacting multi-body system. However, not all nuclei
can be treated with the same success (see $^{40}$Ca) when several
closely spaced neutron hole states can be excited.

However, this reaction seems an ideal tool to study light nuclei with high
precision. Here, mostly nuclei are studied with simple shell
structure of the hypernucleus or the core nucleus ($^4$He, $^6$Li,
$^8$Be, $^{12}$C, $^{16}$O). Achievements of the past have been
first estimates of the \LA~LS-coupling, the \LA-N SS-coupling and
the study of core excitation of Lambda-hypernuclei. Since then the
main aim of many groups is the precise determination of the LS coupling,
SS-strength and the study of hypernuclei using polarization.
\subsection{Determination of the SS-coupling}
The determination of the \LA-N spin-spin coupling is very
difficult as it is expected to be very small. Since usually the
experimental energy resolution of a spectrometer system is in the
order of 2 MeV (close to the size of spin-spin splitting in the NN
case) we may only detect this splitting using $\gamma$-spectroscopy
of \LA-hypernuclei. A good candidate is $^7_{\Lambda}$Li where the
$^6$Li core may be described in the cluster model as $\alpha$+d
(or $^3$He + t, the latter one for higher excitation). We may
obtain a first estimate of the SS-coupling, being typically very
short ranged, assuming the interaction to proceed via 1$^-$
exchange, analog to the 1-photon or 1-gluon exchange. In this case
we may just scale the SS-splitting (0$^+$-1$^+$) of  $^6$Li (~3.56
MeV) with the ratio of magnetic moments of Lambda and nucleons.

\begin{figure}[htbp]

\vspace{-.6 cm} 
\hspace{8 cm}
\begin{minipage}[htpb]{8 cm}

  \begin{center}
    \includegraphics[width=4cm,draft=false]
    {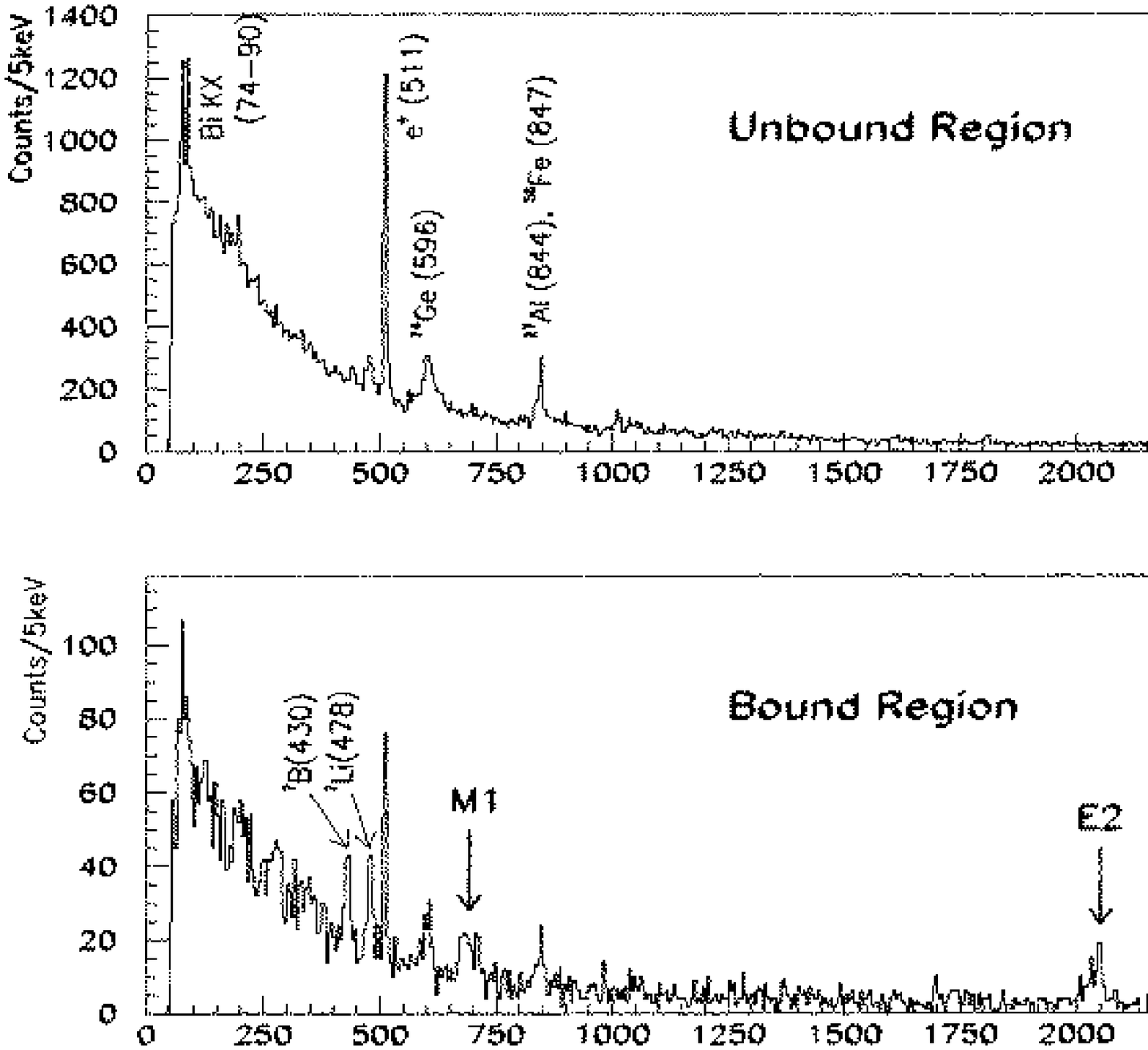}
   \end{center}

   \vspace{-0.9 cm}
   \caption{\small$^7$Li level scheme and $\gamma$-ray spectrum observed}
   \label{SS_coupling}
\end{minipage}
\end{figure}

\noindent
\begin{minipage}[htpb]{7.5cm}
Taking the average of \LA-p and \LA-n interaction (p and n being
parallel in spin) we obtain ($\mu_{\Lambda}\cdot\mu_N \sim$ -0.65 * (2.8 - 1.9) as compared to
$\mu_p\cdot\mu_n\sim$ -1.9 * 2.8) which gives a factor -0.6/-5.5, thus a factor 9. In a
very nice experiment employing a Ge-ball (with BGO Compton shield)
surrounding the hypernuclear production target \cite{SS-splitting}
Tamura et al. have observed two gamma-transitions tagging the
production of bound hypernuclei using the (\PIP,\KP) reaction
(see fig. \ref{SS_coupling}).
They attribute the line at 690 KeV to the M1
transition $\frac{1}{2}-\frac{3}{2}$, thus the SS-coupling. It is about a factor 5 smaller
than in the NN case. Using
a OBE model for the underlying \LA-N interaction this value could
be reproduced theoretically by Motoba et al. This value is also
compatible with the results from light hypernuclei
($^4_{\Lambda}$He and$^4_{\Lambda}$H). The other line at 2.05 MeV
is attributed to the $\frac{5}{2}^+-\frac{1}{2}^+$ transition.

\end{minipage}

\begin{figure}[htbp]

\vspace{-11.2 cm} 

\hspace{8 cm}
\begin{minipage}[htpb]{8 cm}

  \begin{center}
    \includegraphics[width=5.8cm,draft=false]
    {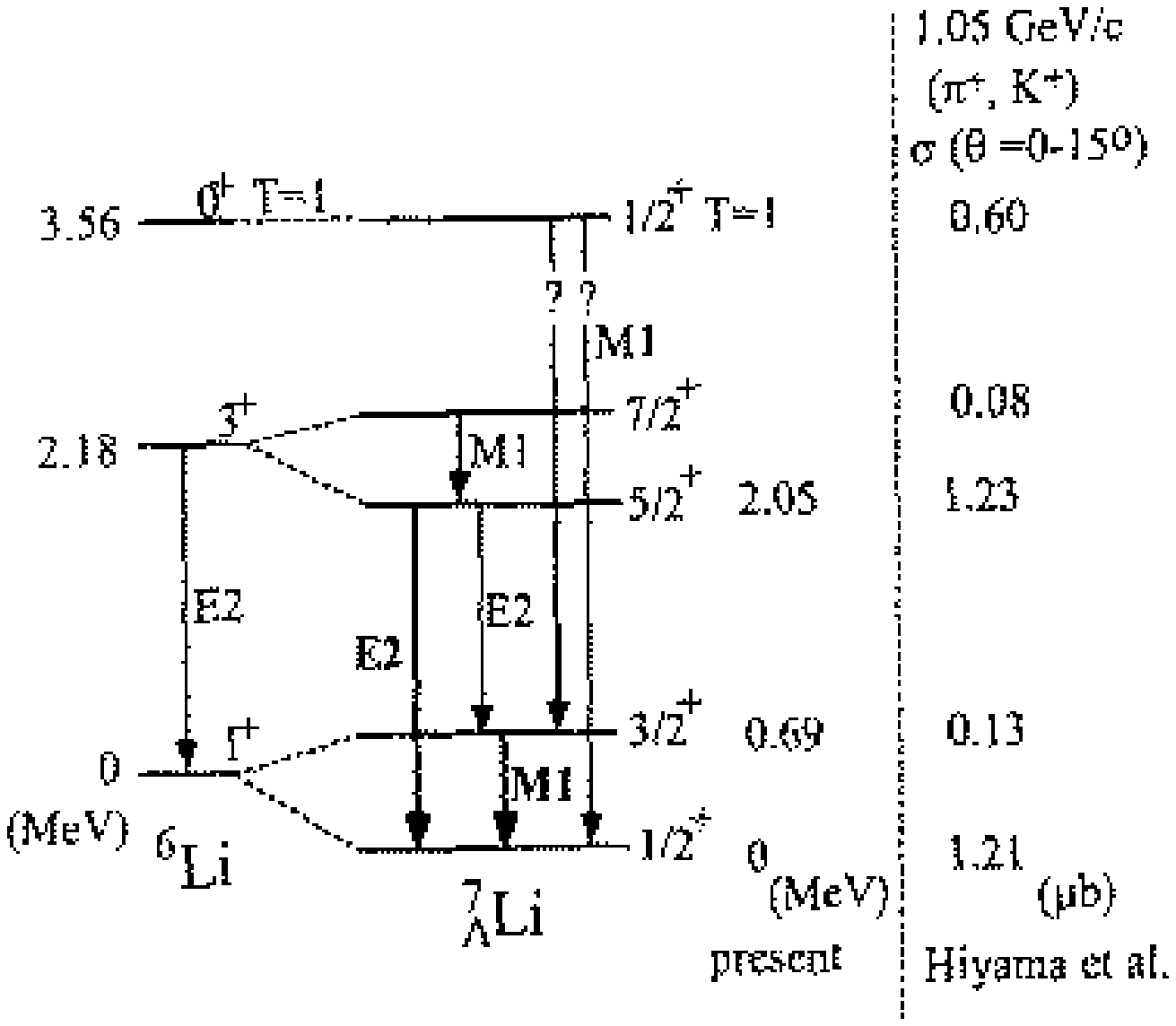}
   \end{center}
   \vspace{-2. cm}
\end{minipage}
\end{figure}

\noindent
\subsection{Determination of the LS-coupling}
\noindent
$^{13}$C is since long considered to be the best
candidate for a determination of the LS coupling of Lambdas. This
state can be seen as a Lambda coupled to a $^{12}$C core. We can
populate the p$_{\Lambda}$ state and try to observe the
$\frac{1}{2}-\frac{3}{2}$ gamma-transition. This is subject of an
ongoing experiment at BNL. In the meanwhile the group at KEK,
again using the (\PIP,\KP) reaction with the superconducting Kaon
spectrometer, has observed the p$_{3/2}$-s$_{1/2}$ splitting to be
9.92 ($\pm$0.13$\pm$0.5) MeV \cite{LS-coupling-1}. Using the

\begin{figure}[htbp]
\vspace{-7.55cm} \hspace{8.5cm}
\begin{minipage}[htpb]{7.5cm}
  \begin{center}
    \includegraphics[width=4cm,draft=false]
    {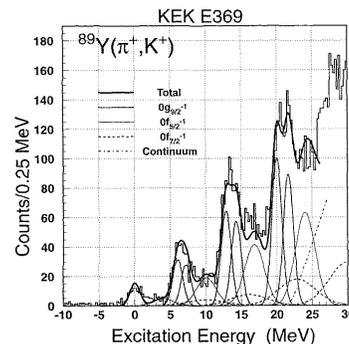}
   \end{center}
   \vspace{-1.25 cm}
   \caption{\small A heavy $\Lambda$-hypernucleus, Yttrium(89)}
   \label{Yttrium}
\end{minipage}
\end{figure}

\noindent
\begin{minipage}[htpb]{7.5cm}
p$_{1/2}$-s$_{1/2}$ gamma transition (10.95$\pm$ 0.1$\pm$ 0.2 MeV)
observed at BNL \cite{LS_BNL} many years ago the LS splitting can be determined
to 1.03 ($\pm$ 0.23 $\pm$ 0.7). This should be compared to an
LS-splitting of 0.72($\pm$ 0.5) MeV observed by the same group in
$^{16}_{\Lambda}$O and previous results from BNL, limiting the LS
splitting to 0.35 ($\pm$ 0.3) MeV \cite{May}.
Using the fact that the LS-coupling increases with L a KEK group
has investigated $^{89}Y_{\Lambda}$ with high resolution allowing
to observe the LS-splitting for the f$_{9/2}$ level (see
fig.\ref{Yttrium})\cite{Yttrium}.
\end{minipage}

\subsection{Lifetimes of hypernuclei}
\LA~in nuclei can decay in two different ways, via mesonic decays
just like a free \LA~(\LA\GOTO p\PIM, n\PIO) or via non-mesonic two
body reactions N\LA\GOTO NN. Thus we expect hypernuclei

\begin{figure}[htbp]
\vspace{-5.8 cm} \hspace{8.5 cm}
\begin{minipage}[htpb]{8 cm}

  \begin{center}
    \includegraphics[width=3.7cm,draft=false]
    {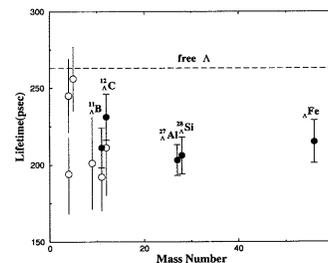}
   \end{center}

   \vspace{-1.3 cm}

   \caption{\small Lifetimes of $\Lambda$-hypernuclei}
   \label{lifetimes_hypernuclei}
\end{minipage}
\end{figure}

\noindent
\begin{minipage}[htpb]{7.5cm}
to live
shorter than free \LA~, depending on the importance of the latter
process. Fig. \ref{lifetimes_hypernuclei} summarizes the presently known
lifetimes of hypernuclei as function of the nuclear mass number
\cite{lifetimes_hypernuclei}. As can be seen, lifetimes drop at
small values of A due to the onset of n.m. decays but stay
constant for nuclei with A$\geq$10. The saturation of n.m. decays
may reflect the saturation of nuclear density and thus the average
number of nucleons seen by the \LA.
\end{minipage}

\vspace{-.2cm}

\subsection{\SI-hypernuclei (are \SI-baryons welcome in nuclei ?)}
\noindent
\begin{minipage}[htpb]{7.5cm}
Since the first observation of \SI-hypernuclei at CERN in Be, C
and O many years ago \cite{Povh_Sigma} much effort has gone into the study of Sigma
hypernuclei. The big puzzle had been the narrow width of the
observed states while theory predicted large widths owing to the
\SI-N $\rightarrow$ \LA-N quenching mechanism. However, until
day, these states could not be observed again neither using
stopped kaons nor the in flight reaction with kinematics similar
to the CERN one by BNL \cite{Sawafta_no_sigma} (E887). 
However, the search for
\end{minipage}

\begin{figure}[htbp]
\vspace{-6.7cm} \hspace{8.5cm}
\begin{minipage}[htpb]{8 cm}
  \begin{center}
    \includegraphics[width=4cm,draft=false]
    {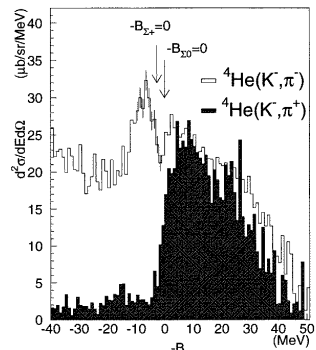}

    \vspace{-1.1 cm}
    \caption{\small $\Sigma$-hypernuclei using helium}
    \label{BNL_Sigma_hypernuclei}
    \end{center}
\end{minipage}
\end{figure}

\newpage
\noindent
a confirmation of  these states finally resulted in the
observation of bound \SI's in He using stopped \KM at KEK. This
measurement had been mentioned time and again since its
publication in 1989.
Recently these states have seen
confirmation using the (\PIP,\KP) reaction at BNL \cite{Sawafta}
at B(\SIO) 4.4 $\pm$ 0.3(stat.) $\pm$ 1(sys) MeV (see fig. \ref{BNL_Sigma_hypernuclei}). The width is
measured to $\Gamma$ = 7$\pm$ 0.7 (stat)$^{+1.2}_{-0}$(sys). This should
be compared to the KEK results obtained with stopped kaons
extracting a binding B$_{\Sigma^+}$ = 2.8$\pm$ 0.7 MeV and
$\Gamma$ = 12.1 $\pm$1.2 MeV \cite{Hayano89}.
Theoretical calculations by Harada et
al.\cite{Harada} give B$_{\Sigma^-}$ = 4.6 MeV and $\Gamma$ = 7.9 MeV using meson
exchange potentials. The non-observation of a similar state in
(\KM,\PIP) producing a \SIM in the nucleus is remarkable. Also a
large isospin dependence of the \SI-N force could be confirmed
since no bound states were observed for \SIM~from the reaction
(\KM,\PIP).
\subsection{Double strange systems (group tourism)}
S=-2 systems have since long attracted attention for two reasons. Ever
since the famous paper of Jaffe \cite{Jaffe} on the existence of a double
strange dibaryon experimentalists have searched for such systems
in various experiments (pp-collisions, heavy ion collisions,
hyperon-collisions). The technique has mostly been to look for
structures in the spectra of invariant mass investigating all
possible final states,  allowing for strong decay to week decay of
the H. Owing to the exotic wave function of this object no
reliable predictions of the production cross section could be
given and thus the non evidence for such a state could not
disprove its existence. More recently the investigations have
focused on the low energy approach, inherited from hypernuclear
physics. Most of the effort is now spent on (\KM,\KP) double
strangeness exchange reaction leaving S=-2 in the target. This
reaction also offers the possibility to look for doubly strange
hypernuclei, a necessary prerequisite for ideas on strange nuclear
matters, so called strangelets. So far, only 3 candidates for
\LA\LA-hypernuclei have been found in emulsion, in part not
uniquely identified, which however set a lower limit to the mass
of the H by the apparent absence of  the fusion mechanism \LA\LA+X
\GOTO H+X of M(\LA\LA)-30 MeV.

\noindent
The following reactions can be investigated:\\\\
\indent
1. \KM $^{12}$C $\rightarrow$ \KP $^{12}$Be$_{\Xi}$\\
\indent
2. \KM $^{12}$C $\rightarrow$ \KP $^{11}$Be$_{\Lambda\Lambda}$ + n
(via intermediate $^{12}$Be$_{\Xi}$ $\rightarrow$
$^{11}$Be$_{\Lambda\Lambda}$ + n)\\
\indent
3. \KM $^{12}$C $\rightarrow$ \KP \XIM X
(\XIM $^{12}$C)atom $\rightarrow$ $^{11}$Be$_{\Lambda\Lambda}$ +
n\\
\indent
4. \KM $^{12}$C $\rightarrow$ \KP \XIM X    (\XIM p($^{12}$C))atom $\rightarrow$ H +
X\\
\indent
5. \KM $^{12}$C $\rightarrow$ H + X\\
\indent
6. \KM p $\rightarrow$ \XIM \KP  \hspace{2.0cm} (\XIM D)atom $\rightarrow$ H +
n\\
\indent
7. \KM $^3$He $\rightarrow$ Hn   \hspace{2.5cm} well calculable\\

\begin{figure}[htbp]
\vspace{-12.5 cm} \hspace{8.5 cm}
\begin{minipage}[htpb]{8 cm}

  \begin{center}
    \includegraphics[width=5.5cm,draft=false]
    {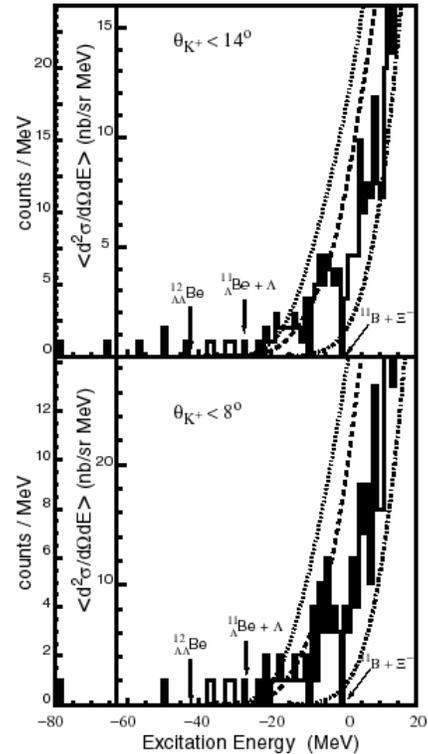}
   \end{center}
   \vspace{-1.5 cm}
   \caption{\small Preliminary data: Missing mass spectrum for a (K$^-$,K$^+$) reaction at BNL}
   \label{ss_hypernuclei}
\end{minipage}
\end{figure}

Fig.\ref{ss_hypernuclei} shows a recent \KP spectrum for such a reaction on nuclei
\cite{Xi_hyp_BNL} converted into bindung energy.
At positive binding energies quasi free
\XIM-production takes place, well reproduced assuming a Fermi-distribution of
the nucleons. At the low end tail (high binding) we expect events with
subthreshold production, namely the production of \XIM's in a
bound nuclear state (reaction 1). If resolution would allow, a
peak at the high end tail (large binding energy for \XIM's) would
give evidence for \XIM-hypernuclei, providing the \XIM-N potential
is strong enough. Possibly some first events of thiskind have been observed.
As for \SI~an exothermic reaction exists in
nuclei, \XIM + p $\rightarrow$ \LA\LA~which is more the probable
process for low binding energies (reaction 2). At low \KP momenta
(large momentum transfers) we produce a \XIM~beam, which
subsequently interacts with the target, slows down, forms
\XIM-atoms and by the strangeness exchange reaction again leads to
2 Lambda which may form a H-dibaryon or double \LA-hypernucleus
(reaction 3 and 4).

\noindent
\begin{minipage}[htpb]{7.5cm}
Several events have been identified for the production of doubly
strange hypernuclei without a clear identification and a
determination of the binding energy. Also production of 2 strange
hypernuclei was observed. The experimental technique employed is a
hybrid system of emulsion and scintillating fibers (SciFi) to link in- and
out-going tracks to the emulsion pictures. At BNL one has employed high
density carbon targets surrounded by SciFi arrays to track the
decay of the strange states. In a new experiment at BNL one uses a
set of cylindrical wire chambers within a solenoid to determine
the momenta of possible decay products of light double strange
hypernuclei produced in a Be target (assuming \LA\LA+core undergo
fission to produce $^6$He$_{\Lambda\Lambda}$ or
$^5$He$_{\Lambda\Lambda}$) with an energy resolution of 1.7 MeV
FWHM. The mass of the double strange hypernucleus can either be inferred
from the \KM spectrum (reaction 1), from the recoil spectrum of n
(reaction 2) or from the spectroscopy of the decay chain of the
\LA\LA-hypernucleus.

\end{minipage}

\noindent
Results:

\begin{enumerate}

\item
For the H-production KEK experiment E224 reported an excess in events in the \LA\LA~invariant
mass spectrum using a scintillating fiber target.

\vspace{-.4cm}
\item No structure could be observed in the neutron energy spectrum from E885 for
$_{\Lambda\Lambda}$Be.

\vspace{-.4cm}
\item E885 has repeated E224 with 10 times higher statistics (no results yet) for the
H-search

\vspace{-.4cm}
\item E885 sees excess of events for large binding energies of \XIM in Carbon using a CH$_2$
target, as compared to calculated cross section. Comparing with
two calculations assuming different V$^0_{\Xi}$ potential shows a
favoring of V$^0_{\Xi}$ ~ -16 MeV.

\vspace{-.4cm}
\item No evidence for structures in the neutron spectrum could be found in experiment E813 (BNL)
using reaction 6.

\vspace{-.4cm}
\item No evidence for structures in n TOF spectrum from reaction 7 at E836
(BNL)

\vspace{-.4cm}
\item
No evidence for structures using $^6$Li target (E836) or $^{12}$C target (E224
KEK)
\end{enumerate}

\section{Future plans:}
In the near future I expect further progress in the field of
charmed baryon spectroscopy, charge radii of hyperons and
hypernuclear physics. In particular the various efforts to extract
the spin-orbit interaction from data to be taken on $^{13}$C is
very promising. Also the role of \XI~in nuclei shall be cleared
during the next years hopefully leading to a first identification
of the ground state. In addition to these topics mentioned already
above other measurements are expected to be done in the context of
the COMPASS experiment at CERN. They address the polarizability of
the kaon and the role of strange quarks in the nucleon on what
concerns the total spin of the nucleon. Continuing efforts at GSI
on bound mesons in nuclei we may expect the study of mass
modifications of the kaon in nuclear matter, in analogy to the
recent experiment with bound pions using heavy nuclei, an effect
related to the restoration of chiral symmetry.

In the more distant future we may think of implanting charmed
hadrons into nuclei, using new facilities currently under
discussion (like a \PBAR~storage ring at GSI).

\section{Conclusions}
We have seen that the reaction of a well known system to a strange
or heavy intruder shows many similarities for hadrons and nuclei,
even if the underlying physics may be very different. This is true
concerning the single particle picture, the masking of particular
parts of the usually very complex interaction (e.g. spin-spin
(hadron) or spin-orbit interaction (nucleus)) or the modification
of the properties of the 'tourist' like its lifetime which can be
caused by additional reaction channels only possible in the new
surrounding.

These effects can now be studied in detail owing to new and
precise data coming from many laboratories around the world, both
in high energy physics and in nuclear physics.

The similarities of the phenomena observed in the two systems is
remarkable.

\end{document}